# Nanoscale Determination of the Mass Enhancement Factor in the Lightly-Doped Bulk Insulator Lead Selenide

Ilija Zeljkovic[1], Kane L Scipioni[1], Daniel Walkup[1], Yoshinori Okada[1,2], Wenwen Zhou[1], R. Sankar[3], Guoqing Chang[4], Yung Jui Wang[5], Hsin Lin[4], Arun Bansil[5], Fangcheng Chou[3], Ziqiang Wang[1] and Vidya Madhavan[1,6,*]

[1]*Department of Physics, Boston College, Chestnut Hill, Massachusetts 02467, USA* [2]*WPI-AIMR, Tohoku University, Sendai, 980-8577, Japan* [3]*Center for Condensed Matter Sciences, National Taiwan University, Taipei 10617, Taiwan* [4]*Graphene Research Centre and Department of Physics, National University of Singapore, Singapore 117542* [5]*Department of Physics, Northeastern University, Boston, Massachusetts 02115, USA* [6]*Department of Physics and Frederick Seitz Materials Research Laboratory, University of Illinois Urbana-Champaign, Urbana, Illinois 61801, USA*

[*]Corresponding author: vm1@illinois.edu

## Abstract

Bismuth chalcogenides and lead telluride/selenide alloys exhibit exceptional thermoelectric properties which could be harnessed for power generation and device applications. Since phonons play a significant role in achieving these desired properties, quantifying the interaction between phonons and electrons, which is encoded in the Eliashberg function of a material, is of immense importance. However, its precise extraction has in part been limited due to the lack of local experimental probes. Here we construct a method to directly extract the Eliashberg function using Landau level spectroscopy, and demonstrate its applicability to lightly-doped thermoelectric bulk insulator PbSe. In addition to its high energy resolution only limited by thermal broadening, this novel experimental method could be used to detect variations in mass enhancement factor at the nanoscale. As such, it opens up a new pathway for investigating the effects of chemical defects, surface doping and strain on the mass enhancement factor.

## Introduction

Interactions between electrons and their surroundings within a solid is ubiquitous in systems ranging from simple metals [1], to more exotic materials such as graphene [2,3], high-temperature superconductors [4–9] and topological insulators [10–16]. Of particular interest is the coupling between electrons and phonons. In general, phonons accompany the movement of free electrons, which in turn gives rise to quasiparticles of decreased mobility and increased effective mass. This process is theoretically described by the Eliashberg function $α^2F(ω;k,E)$ (Ref. [17]), which captures the scattering of quasiparticles by coupling to a bosonic mode of frequency ω. Although not directly measurable by experiments, the Eliashberg function is related to the complex self-energy of the system ($Σ(k,E)$) (Ref. [17]), an experimentally-accessible quantity which encodes both the quasiparticle energy ($Σ'(k,E)=Re\{Σ(k,E)\}$) and its lifetime ($Σ''(k,E) =Im\{Σ(k,E)\}$). $Σ'(k,E)$ and $Σ''(k,E)$ are related to one another by Kramers-Kronig relations, so knowing only one of them should, in principle, be sufficient to determine the total self-energy $Σ(k,E)$, and thus the Eliashberg function itself [18]. Since many fundamental properties, such as the electron mass enhancement factor ($λ$), can be calculated from the Eliashberg function, its precise experimental determination is of immense importance.

In topological materials, there is mounting evidence for the role of phonons in core physical processes [10,19]. Although relatively well-described within the framework of non-interacting Dirac theory, surface state Dirac fermions in topological materials are experimentally found to interact with lattice excitations [10–16], leading to the renormalization of the underlying surface state band structure. The role and magnitude of this interaction are still under debate, as vastly different results have been obtained, not only across different techniques, but also utilizing the same experimental probe on nominally identical compounds [13,14]. For example, the measured $λ$ in a typical three-dimensional topological insulator $Bi_2Se_3$ is found to range from as low as $λ$ ~0.17 [13] (measured at 20 Kelvin) to $λ$~3 [14] (measured at 7 Kelvin) in recent laser-based angle-resolved photoemission spectroscopy (APRES) studies reported by two different groups. Careful analysis by Howard and El-Batanouny showed that different acquisition temperatures (7 Kelvin vs. 20 Kelvin) would account for some of the discrepancy, but only less than a factor of 2 between the two studies [20]. Other possible reasons for this discrepancy may be due to the resolution of the technique, or the inability to detect the 3 meV phonon peak in Ref. [13] which was reported by Kondo *et al.* [14]. Another contributing factor might lie in the intrinsic structural inhomogeneity (local chemical potential variation, step-edges, defects, strain/stress, etc.) present on the surface of these materials, either as an inevitable consequence of the cleavage process or the varying sample quality. Thus, developing a high-resolution technique for local determination of electron-phonon coupling (EPC) on nanometer length scales with simultaneous imaging of the surface and the local density of states would be of tremendous importance, not only for obtaining $\lambda$ with a measure of reliability, but also for a deeper understanding of the factors affecting EPC, and ultimately for measuring the Eliashberg function for interacting systems such as superconductors or charge density wave materials where the EPC plays an important role in determining their ground state.

To illustrate subtle self-energy effects present due to the interaction of phonons and electrons, let us use a single electronic band crossing the Fermi level as an example (Fig. 1). In the absence of EPC, this band is expected to smoothly cross the Fermi level. However, when EPC is turned on, an antisymmetric kink develops in the dispersion near the Fermi level, depleteing the states below and populating the states above the Fermi level. To reliably extract this band renormalization, high-resolution experimental determination of the electronic band structure is crucial. Scanning tunneling microscopy (STM) has been successfully applied to nanoscale band structure mapping utilizing either the method of quasiparticle interference (QPI) [21,22], or Landau level (LL) spectroscopy [23–25]. In relatively simple systems such as Ag(111), QPI imaging has proven effective for the detection of small kinks in the electronic dispersion, which allowed the approximation of the electron mass enhancement $\lambda$ [26]. Applying the same technique to Fe-based high-temperature superconductorts has revealed the momentum-space anisotropy in self-energy likely originating from antiferromagnetic spin fluctuations [9]. Therefore, this method could naturally be extended to other materials. However, it cannot be used in many three-dimensional topological materials as the fundamental nature of the topological surface states prohibits quasiparticle backscattering [27], and therefore prevents the observation of QPI in the desired energy range [28,29]. Furthermore, the extracted band structure from QPI measurements is often noisier [25], requires longer time acquistion, and provides lower energy resolution compared to the band mapping using LL spectroscopy. In the following paragraphs, we will describe the first utilization of LL band mapping to exctract the self-energy, the Eliashberg function and $\lambda$ in a trivial bulk insultor $Pb_{1-x}Sn_xSe$ with $x$~0.09.

## Results

**Surface characterization.** Single crystals of $Pb_{1-x}Sn_xSe$ are prepared using the self-selecting vapor growth method, cleaved at ~77 Kelvin in ultra-high-vacuum, and immediately inserted into the STM where they are held at 4 Kelvin during data acquisition. Although the process of cleaving can result in mutiple surface step-edges and even distinct cleavage planes, we are able to find microscopically large, atomically flat regions characterized to be the (001) surface based on the square atomic lattice observed in STM topographs (Fig. 2(a)). The average conductance d$I$/d$V$ spectrum, which is directly proportional to the local density of states, exhibits a clear gap of ~50 meV (Fig. 2(b)), confirming that this composition would be a band insulator without the intrinsic doping due to defects which moves the Fermi energy into the conduction band. Although $Pb_{1-x}Sn_xSe$ is non-topological at the alloying composition studied, it hosts distinct trivial surface states which can be observed and studied using STM [30]. We note here that the gap measured by us sets a lower limit for the band gap; the actual value may be larger but masked due to the surface state. We proceed to describe the utilization of LL spectroscopy to map an electronic band crossing the Fermi level for the purposes of quantifying many-body interactions.

**Electronic band structure mapping using Landau level spectroscopy.** When a magetic field is applied perpendicular to the surface of a sample, LLs emerge as peaks in d$I$/d$V$ spectra on top of the U-shaped background, which signal the quantization of the two-dimensional surface states (Fig. 3(a,b)). Using semi-classical approximation, it can be shown that for an isotropic Fermi surface, the magntude of the momentum vector ($k$) at any given energy $E$ is proportional to $(NB)^{0.5}$, where $N$ is the LL index, $B$ is the applied magnetic field, and $E$ is the energy of the LL peak. For a non-isotropic Fermi surface, the Fermi surface area is proportional to $NB$, which can be mapped to an effective (angle integrated average) momentum given by $(NB)^{0.5}$. In either case, this allows us to use LLs to measure the angle integrated dispersion with high precision.

Our first task is to plot the LL energy as a function of $(NB)^{0.5}$. Since precise determination of LL peak positions is of extreme importance for our study, we normalize each d$I$/d$V$ spectrum acquired in a magnetic field (Fig. 3(b)) before fitting a Lorentizan curve to locate the postion of each LL peak (see Supplementary Figure 1 and Supplementary Note 1). We note that this process is only possible if all d$I$/d$V$ spectra are acquired over the *identical* region of the sample with *no tip changes*. Next, we index the LLs with integer numbers as shown in Fig. 3(b), and plot their positions as a function of average integrated momentum to obtain a continuous dispersion (Fig. 3(c)) (Supplementary Figure 2). Since the data was acquired in small magnetic field increments of 0.1 T, we are able to obtain nearly 100 data points within the narrow energy range of +/- 30 meV around the Fermi level. The energy resolution of our experiments is only limited by thermal broadening (~$4k_BT$, where $k_B$ is the Boltzmann constant and $T$ is the temperature) and is ~1.5 meV at 4 Kelvin.

**Extraction of the Eliashberg function and mass enhancement factor.** The real part of the self-energy of the electrons $\Sigma'(k,E)$ can be calculated as the difference between the measured dispersion $\varepsilon(k)$ shown in Fig. 3(c) and the bare quasiparticle dispersion $\varepsilon_0(k)$ (Fig. 4(a)). To get a physical intuition on the bare quasiparticle dispersion, we perform tight-binding calculations on $x$~0.09 $Pb_{1-x}Sn_xSe$ to obtain the theoretically predicted SS dispersion. As expected, this dispersion is nearly linear with a very small quadratic correction. It can be approximated as a second degree polynomial in the form of $\varepsilon_0(k) = \alpha(k-k_f) + \beta(k-k_f)^2$, where $k_f$ is the momentum wavevector at the Fermi level (Supplementary Note 2). Using coefficients $\alpha$ and $\beta$ (adjusted by less than 6% to match our experimental data at energies away from the Fermi level), we can determine $\Sigma'(k,E)$ (Fig. 4(a)). The extracted $\Sigma'(k,E)$ is anti-symmetric with respect to the Fermi level, with positive values for the occupied states and negative values for the unoccupied states. $\Sigma'(k,E)$ and the Eliashberg function $\alpha^2F(\omega)$ are related by the integral equation: $\Sigma'(k,E) = \int_0^\infty d\omega\ \alpha^2F(\omega)\ K[E/(kT),\ \omega/(kT)]$, where $K[y,\ y'] = \int_{-\infty}^\infty dx\ f(x-y)\ 2y'/(x^2-y'^2)$ and $f(x)$ is the Fermi distribution function. The inversion of this integral to calculate $\alpha^2F(\omega)$ is mathematically unstable and results in unphysical solutions [18]. However, this problem can be circumvented by the application of the maximum entropy method (MEM) (Ref. [18]) to fit $\Sigma'(k,E)$ and calculate the $\alpha^2F(\omega)$ of our system (Fig. 4(b)) (Supplementary Notes 3 and 4). In contrast to the use of this method on real self-energy obtained from ARPES [18,31], our measurements allow access to *both* occupied and unoccupied states, and we can use the whole energy range of these to calculate $\alpha^2F(\omega)$, thus utilizing more information to converge to the correct $\alpha^2F(\omega)$. The mass enhancement factor $\lambda$ is easily calculated from $\alpha^2F(\omega)$ as $\lambda = 2 \int_0^\infty d\omega\, \alpha^2F(\omega)/\omega$. From our data in Fig. 4(a), we get $\lambda$ to be 0.11 ± 0.05. Our measurements reveal weak electron-phonon coupling in this material, which may play a role in the high thermoelectric power observed by other techniques [32]. Interestingly, we note that surface $\lambda$ measured in our experiments is comparable to the bulk $\lambda$ reported on similar compounds obtained using Shubnikov-de Haas oscillations [33].

To demonstrate the validity of our method, we perform the following tests. Both $\alpha^2F(\omega)$ and $\lambda$ determined by separately using the dispersion of the occupied electronic states ($\alpha^2F_-(\omega); \lambda_-$) and the unoccupied states ($\alpha^2F_+(\omega); \lambda_+$) are expected to yield identical results. To check this, we first calculate and directly compare $\alpha^2F_-(\omega)$ and $\alpha^2F_+(\omega)$ functions. Remarkably, these two provide an excellent match within our experimental resolution as the principle phonon modes in both are at nearly identical energies (Supplementary Figure 3), which further confirms the high quality of our experimental data. Additionally, $\lambda_-$ and $\lambda_+$ extracted from curves in Supplementary Figure 3 are the same within the measurement error.

Finally, we apply our method to a second system, $Bi_2Se_3$, in order to search for the expected signatures of electron-phonon coupling (Supplementary Note 5). Although the reported magnitudes of $\lambda$ in $Bi_2Se_3$ vary widely across different experiments as discussed in the introductory paragraphs, even the smallest detected $\lambda$ (~0.17 in Ref. [13]) should in principle result in band renormalization large enough to be detected by our novel experimental method. In this endeavor, we use the data from one of the pioneering LL studies in $Bi_2Se_3$ acquired by Cheng *et al.* [34]. The extracted SS band dispersion shows a prominent deviation from the smooth evolution noticeable near the Fermi level which clearly signals the presence of EPC (Supplementary Figure 4(a)). Furthermore, by calculating the real part of the self-energy (Supplementary Figure 4(b)) we determine $\lambda$ to be ~0.26, which is within the range of reported values on this material. We note that although the analysis presented here on Bi$_2$Se$_3$ suggests EPC weaker than some of the other studies [14–16], we cannot exclude the presence of ~3 meV peak reported by Kondo *et al.* [14] or ~8 meV peak reported by Sobota *et al.* [16], and higher resolution future studies may be needed. Nevertheless, the available data and analysis presented here present yet another confirmation of the validity of our methodology.

## Discussion

We expect that this newly developed method of determining EPC by Landau level spectroscopy will be especially useful for two dimensional systems like graphene or ultra thin films, as well as in materials where surface states play an important role in determining the electronic properties. In many topological insulators, bulk and surface state electronic bands occur close in energy and disentangling the two using ARPES can be challenging. Our method provides a natural solution to this problem as the strongly $k_z$-dispersing bulk bands would not be observed using STM, which allows us to examine the effect of phonons exclusively on the topological surface states. Furthermore, phonon modes and EPC itself are likely sensitive to the presence of inhomogeneous structural distortions, and our work provides a way to probe this. For example, 'stripe'-like inhomogeneity in Bi$_2$Te$_3$ single crystals [28] and surface dislocations in Bi$_2$Se$_3$ thin-films [35] have already shown to result in a prominent change in the local density-of-states, but the evidence of any resulting modification in the phonon spectrum is still lacking. Our new method could provide further insight into the electron-phonon dynamics in these materials at the nanoscale, and could facilitate engineering other novel systems with more desirable properties.

# Figures

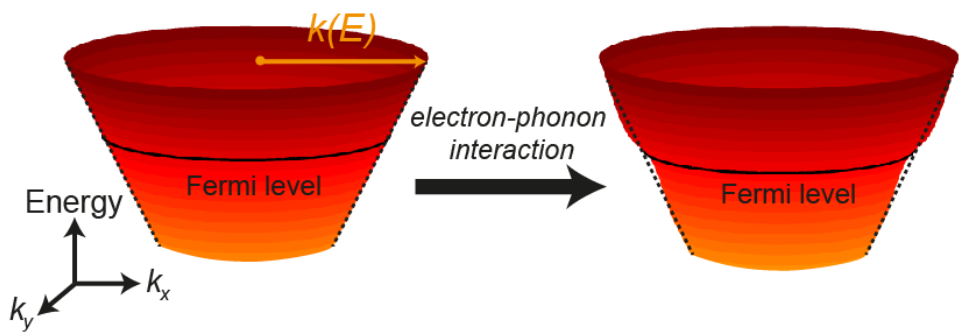

**Figure 1.** The schematic of the electron-phonon coupling (EPC) effect on the band structure of topological materials. In the absence of EPC (left), the Dirac SS band smoothly crosses the Fermi level. However, when EPC is turned on (right), electronic density of states gets renormalized, resulting in a "kink" in the SS dispersion near the Fermi level.

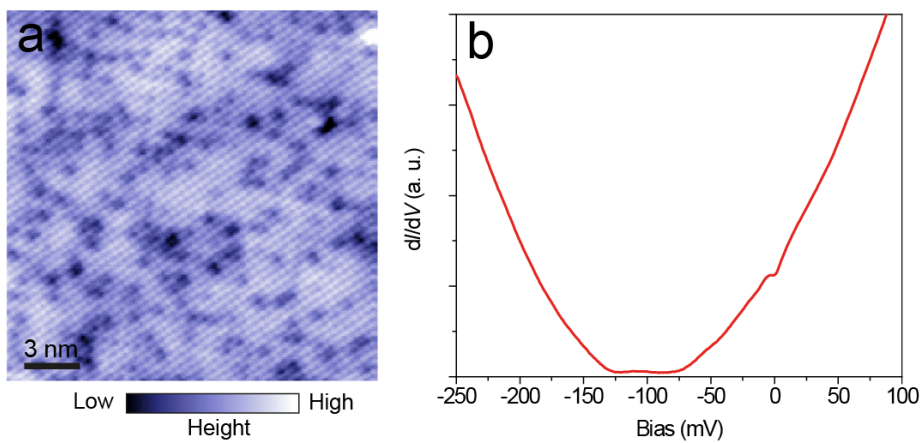

**Figure 2.** Typical STM topograph and d$I$/d$V$ spectrum. (a) Typical STM topograph of the (001) surface showing a clear square lattice. (b) Average d$I$/d$V$ spectrum obtained over the region of the sample in (a), showing an integrated density of states gap of ~50 meV which sets a minimum value for the band gap at this composition. Setup conditions in: (a) $I_{set}$=10 pA, $V_{set}$=-100mV; (b) $I_{set}$=100pA, $V_{set}$=-100mV.

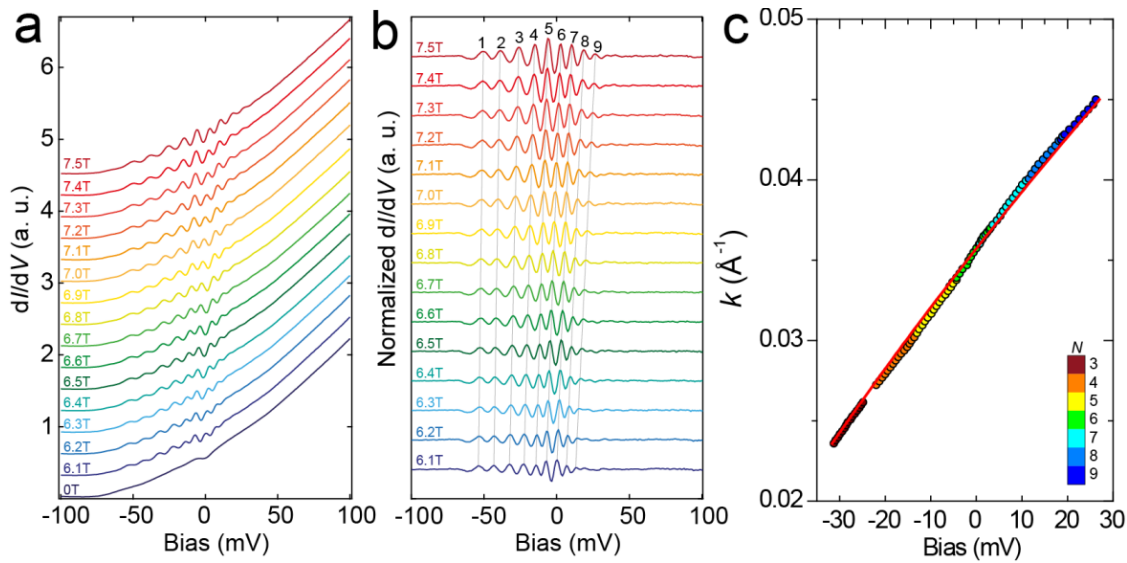

**Figure 3.** Landau level spectroscopy and the extraction of electronic band dispersion. (a) Average d$I$/d$V$ spectra, vertically offset for clarity, obtained at differnet applied magnetic fields. Each spectrum is determined by averaging 64 point-spectra, spaced equally along a 20 nm linecut across the region of the sample shown in Fig. 2(a). (b) Normalized d$I$/d$V$ spectra, vertically offset for clarity (see Supplementary Note 1 for normalization details). (c) Average integrated momentum $k$ vs. bias dispersion of a single SS band crossing the Fermi level (circles), and the bare quasiparticle dispersion (red) (Supplementary Note 2). Different colors of circles indicate different Landau level indices. Setup condition in (a) are: $I_{set}$=100pA, $V_{set}$=-100mV.

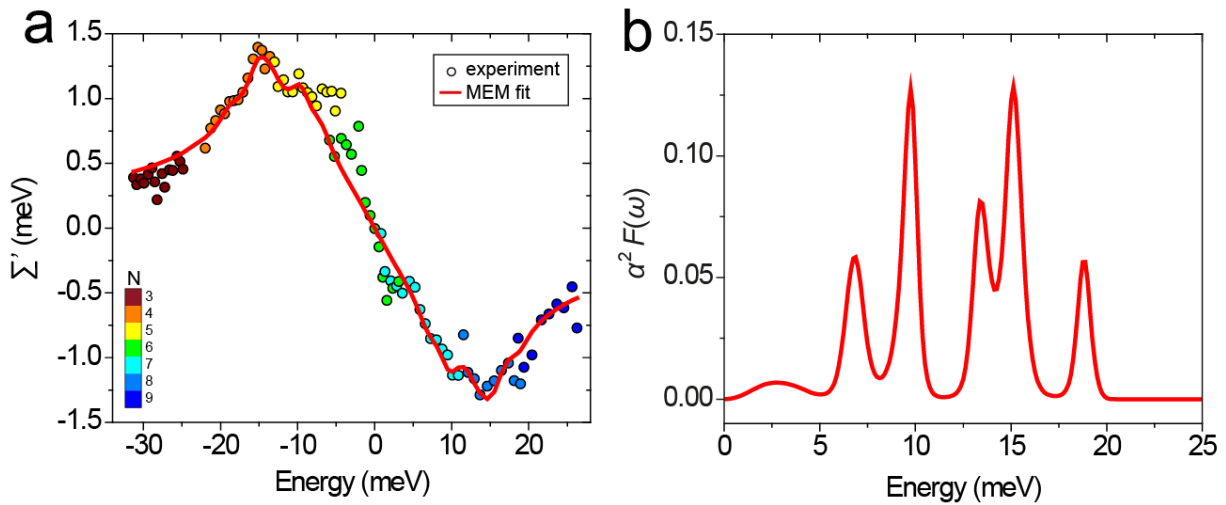

**Figure 4.** Real part of the self-energy ($\Sigma'$) and the extraction of the Elaishberg function. (a) Experimentally determined $\Sigma'$ (open circles) and the maximum entropy method (MEM) fit (red line). Colors of circles indicate the index of the Landau level used. (b) The extracted Elaishberg function (red) obtained by using the experimental data in (a).

# Supplementary Material

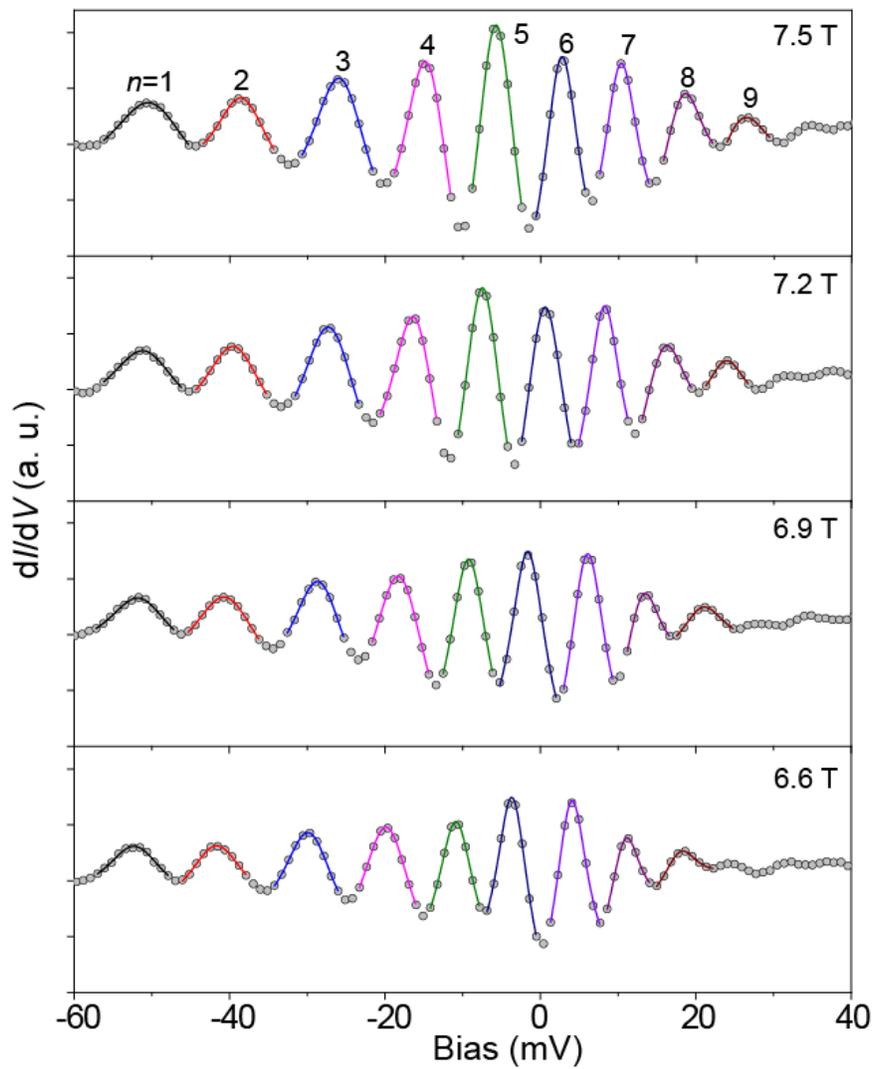

**Supplementary Figure 1.** Normalized d$I$/d$V$ spectra (gray circles), and Lorentzian fits to each Landau level (curves) for several representative magnetic fields.

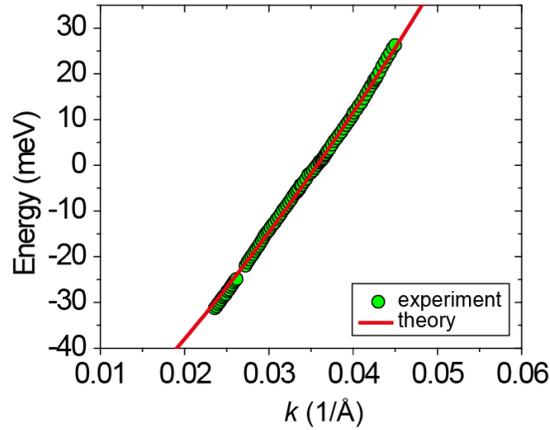

**Supplementary Figure 2.** Comparison of experimental data (green circles) and the theoretically obtained dispersion (red line). Fermi level for the theoretical dispersion was set so that $k_f$ of the experiment matches the corresponding $k_f$ in theory. As it can be seen, an excellent match is obtained.

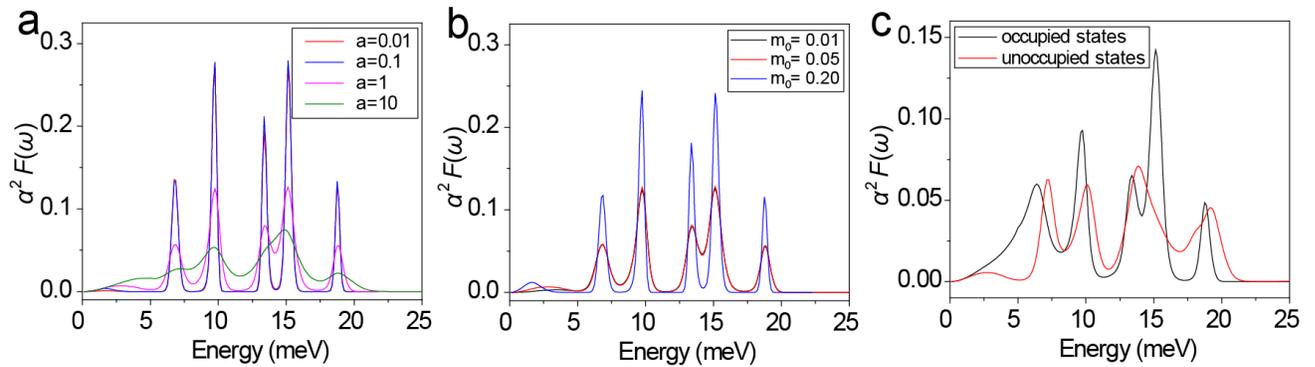

**Supplementary Figure 3.** Robustness of MEM algorithm. Comparison of extracted Eliashberg functions for different parameters (a) $a$ and (b) $m_0$, showing no qualitative difference. (c) Extracted Eliashberg functions if only occupied or unoccupied states are considered, which show an excellent agreement.

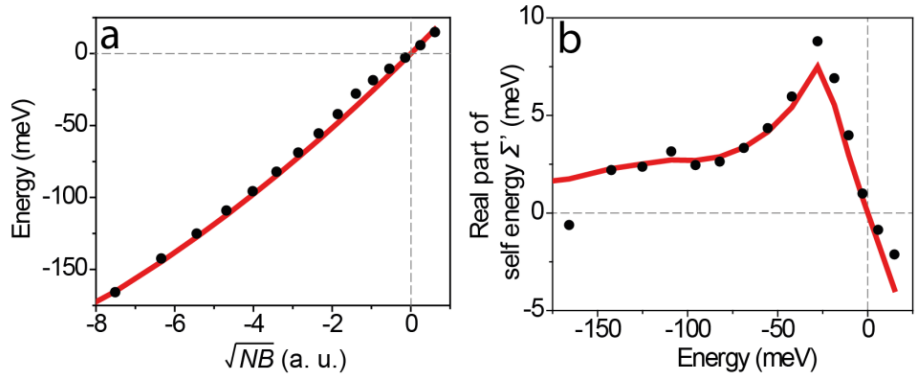

**Supplementary Figure 4.** Extracting the real part of the self-energy in $Bi_2Se_3$. (a) Energy vs. sqrt(NB), where N is the LL index and B is the magnetic field (8 T for the dispersion shown). The dispersion was obtained by digitizing and fitting the raw LL spectroscopy data from Reference 1. Red line represents bare quasiparticle dispersion in the absence of electron-phonon coupling. (b) Experimentally determined real part of the self-energy Σ' (circles) and the maximum entropy method (MEM) fit (red line).

| Experimental surface phonons (meV) | 7.0 | 9.8 | 13.3 | 15.2 | 18.9 |
|---|---|---|---|---|---|
| Theoretical bulk phonons (meV) [2] | 5.6 | 10.2 | 14.0 | 15.8 | 17.8 |

**Supplementary Table 1.** Comparison of the main modes obtained from MEM fitting in our lightly-doped PbSe sample and calculated phonons from first-principles calculations on pure PbSe [2].

## Supplementary Note 1. d*I*/d*V* normalization and Landau level peak fitting

The normalization of d*I*/d*V* spectra involves two steps. First, we subtract the 0 T average d*I*/d*V* spectrum from the spectra obtained over the same area of the sample in magnetic field. Then, we subtract a third degree polynomial to offset for the varying amounts of thermal and piezo drifts during the acquisition of different data sets.

We fit a simple Lorentizan curve to each Landau level to determine its position. As it can be seen for several representative curves (Supplementary Figure 1), we get an excellent match between the Lorentizan fits and the experimental data.

## Supplementary Note 2. Determination of the electronic band structure

### Experimental extraction

Having determined the positions of each Landau level peak, we utilize the semi-classical approach to calculate the area of the constant-energy contours (CECs) in momentum space as:

$$S_n = \frac{2\pi e}{\hbar} nB \quad (1)$$

where *n* is the LL index, and *B* is the applied magnetic field. By approximating the CECs as circles with radius $k(E)$, we get the average integrated momentum dispersion as $k(E) = \sqrt{\frac{S_n}{\pi}}$, which is plotted in Fig. 3(c) of the main text. Experimentally determined $k_f \equiv k(E_f)$=0.036 Å$^{-1}$.

### Theoretical band structure calculations

The bulk band structures of PbSe and SnSe in an ideal rock-salt crystal are computed within the framework of the density functional theory (DFT) using projector augmented wave method as implemented in the VASP package [3]. The generalized gradient approximation (GGA) (Ref. [4]) is used to model exchange-correlation effects. The cutoff energy 260 eV is used, and the spin orbit coupling (SOC) is included in the self-consistent cycles. The lattice constant of PbSe has been chosen to be 6.31 Å to give a 150 meV band gap to match the recent experimental observation [5]. The lattice constant of SnSe has been chosen to be 6.03 Å to give a band inversion with a band gap value of 600 meV at the L-point. Tight-binding parameters based on Wannier functions for both PbSe and SnSe are obtained by wannier90 (Ref. [6]) with a basis set consist of Pb/Sn $p_x$, $p_y$, $p_z$ and Se $p_x$, $p_y$, $p_z$. The tight-binding parameters for $Pb_{1-x}Sn_xSe$ are obtained by linear interpolation between two end compounds for selected Sn composition of x~0.09. These parameters are then used for a slab of 200 atomic layers to calculate the CECs of the surface states. Theoretically calculated average momentum dispersion is then:

$$E - E_f = \alpha(k - k_f)^2 + \beta(k - k_f) \quad (2)$$

where $= 14.24\ eV Å^{-2}$, $\beta = 2.64\ eV Å^{-1}$ and $k_f = 0.036\ Å^{-1}$. Although this dispersion provides an excellent visual match with the data (Supplementary Figure 2), it also slightly deviates from the experimental data points at energies away from the Fermi level and results in unphysical, negative real-part of the self-energy of the system. By adjusting $\beta$ by ~6% and leaving $\alpha$ unchanged, we circumvent this problem and proceed to use these adjusted parameters for the bare quasiparticle dispersion in MEM algorithm in the main text.

## Supplementary Note 3. Robustness of the MEM algorithm

The code used in this section is a modified version of the program developed by Shi et al. [7] which was primarily designed for analyzing ARPES data. We adapted this code to use the information for both the occupied and occupied states which STM has access to. The basic premise of the maximum entropy method (MEM) based inversion of the integral $\Sigma'(k, E) = \int_0^\infty d\omega \alpha^2 F(\omega) K[\frac{E}{kT}, \frac{\omega}{kT}]$ (Supplementary Equation 3) is to maximize the function $L = aS - \frac{\chi^2}{2}$, where $S$ is the Shannon-Jaynes entropy, $\chi^2$ is the standard chi-squared deviations between the raw $\Sigma'$ and the calculated $\Sigma'$, and $a$ is a maximization parameter that is optimally determined in the manner described elsewhere [7]. In this section, we show that the qualitative results of the algorithm are relatively unaffected for a wide range of parameters used.

The Eliashberg function $\alpha^2F(\omega)$ obtained by the process of MEM-based inversion using a manually set parameter $a$ at different values spanning four orders of magnitude is shown in Supplementary Figure 3(a). This comparison clearly shows that the positions of the peaks are well-resolved below $a = 10$, and virtually unchanged for all lower values of $a$. Qualitatively, $a$ is a fitting parameter that determines the extent to which $\alpha^2F(\omega)$ deviates from the initial constraint function during the fitting process, which in our case is mostly a constant value through the range of $\omega$. The main iterative feature in the fitting procedure is a step in the function $L$ towards the global maximum. These steps are translated to a correction in the $\alpha^2F(\omega)$ which is then applied. The function $L$ is recalculated using the new $\alpha^2F(\omega)$, and the process is repeated until the correction in $\alpha^2F(\omega)$ is very small. The size of these corrections is inversely dependent on the parameter $a$, which explains the quantitative difference in the curves shown in Supplementary Figure 3(a). However, the main phonon energies at which the peaks are located are preserved over four orders of magnitude of $a$. We proceed to use the so-called "classical" method of optimization determined by the code provided in Ref. [7] for which $a = 1.010$.

For the constraint function $m(\omega)$, one needs to properly choose, based on *a priori* knowledge, the functional form, and the order of magnitude to obtain a proper physical description of $\alpha^2F(\omega)$. Specifically, the form we used is:

$$m(\omega) = \begin{cases} m_o \left(\frac{\omega}{\omega_D}\right)^2, & \omega_D > \omega > 0 \\ m_o, & \omega_M > \omega > \omega_D \\ 0, & \omega > \omega_M \end{cases} \quad (4)$$

where $m_o$ is a constant set approximately at the average value of the expected $\alpha^2F(\omega)$ and $\omega_M$ is the maximal phonon energy. This is the same form used in Ref. [7] in which they expound upon the motivation. In this study, we choose the maximal phonon energy of 25meV and $\omega_D$ of 5meV. The extracted $\alpha^2F(\omega)$ using different values of $m_o$ is shown in Supplementary Figure 2(b) (we choose $m_o = 0.05$ in the main text). We show that this value does not need to be set to a precise optimum as it can be varied by small amounts about $m_o = 0.05$, without changing the locations of the peaks and their relative intensities.

We move on to the discussion of the errors bars obtained from the MEM fitting. Before the fitting can be initiated, we must be able to calculate

$$\chi^2 = \sum_0^N \frac{1}{N(N-1)} \frac{(Re[\Sigma_N] - Re[\Sigma_{N,fit}])^2}{\sigma_N^2} \quad (5)$$

where the fit to the real self-energy $Re[\Sigma_{fit}]$ is initially obtained from performing the integration of Supplementary Equation 3 using $\alpha^2 F(\omega) = m(\omega)$. However, the values $\sigma_N$ need to be determined in a systematic way that reflects the uncertainty of the raw data points in the real part of the self-energy $Re[\Sigma_N]$. In this calculation, we used a series of quadratic polynomial fits expanded about each data point to calculate the error for each data point. Within +/-5 meV of the Fermi level, we manually assigned an error of +/- 1meV due to a known experimental uncertainty within that range.

One of the main advantages of using LL spectroscopy to map the electronic band dispersion is the ability to independently compare $\alpha^2 F(\omega)$ extracted using exclusively the raw data points for the occupied states, and those describing the unoccupied states. This comparison is shown in Supplementary Figure 3(c) and the two exhibit an excellent agreement. Furthermore, the observed peaks are indeed nearly symmetric around the Fermi level as expected. We note that the result from using the occupied states in the system has sharper peaks than the result obtained by using exclusively the unoccupied states. This is because the calculated errors for the raw data are smaller for the occupied than for the unoccupied states. Furthermore, the data seem to contain more "noise" above the Fermi level, as higher LLs become less prominent and the error in determining their exact position increases. When the errors are larger, a single point's contribution to $\chi^2$ is weighted less because there is inherently less information about the real self-energy. This results in a broader peaks and thus less information about the structure of $\alpha^2 F(\omega)$.

## Supplementary Note 4. Comparison of experimental SS modes and calculated bulk phonons

Supplementary Table 1 shows the comparison between the principal modes obtained from MEM fitting in Fig. 4 of the main text and the bulk phonon modes of PbSe calculated using first-principles method in Supplementary Reference 2.

## Supplementary Note 5. Extraction of EPC constant λ in a three-dimensional topological insulator $Bi_2Se_3$

To determine the EPC constant λ in a three-dimensional topological insulator $Bi_2Se_3$, we use the LL spectroscopy data published in Reference 1. Similarly to the procedure used in the main text, we index the LLs with consecutive integers starting from 0, and plot the energy dispersion as a function of $(NB)^{0.5}$ (proportional to the average angle-integrated momentum k), where N is the LL index and B is the magnetic field (circles in Supplementary Figure 4(a)). To obtain the bare quasiparticle dispersion in this material, we fit a second degree polynomial to raw data, including points away from the Fermi level and the data point at $k_f$ (red line in Supplementary Figure 4(a)). A prominent deviation of experimental dispersion from the bare quasiparticle dispersion is observed near the Fermi level, which is even more obvious in the calculated real part of the self-energy (Supplementary Figure 4(b)). Although the quality of the published experimental data does not allow us to extract all the phonon modes as we were able to do with our own data on lightly doped PbSe, we can calculate λ to be ~0.26, which is within the range of reported values on this material. We repeated the same procedure on multiple magnetic fields in Reference 1 to obtain qualitatively similar results for the self-energy Σ' in Supplementary Figure 4(b) and for λ.

## Supplementary References